\documentclass{article} 
\usepackage{iclr2024_conference,times}

\usepackage{hyperref}
\usepackage{url}

\usepackage{mathtools,amssymb,amsthm,bm,bbm}
\usepackage[Symbol]{upgreek}
\usepackage{graphicx} 
\usepackage{algorithm,algpseudocode}
\usepackage{booktabs}

\def\Nd{\mathbb{N}}

\def\Rd{\mathbb{R}}
\def\Ed{\mathbb{E}}

\def\Nc{\mathcal{N}}

\def\rbr#1{\left(#1\right)}
\def\sbr#1{\left[#1\right]}
\def\cbr#1{\left\{#1\right\}}

\def\inh{\text{inh}}
\def\exc{\text{exc}}
\def\no{\text{no}}
\def\GSoftmax{\operatorname{Gumbel-Softmax}}

\title{One-hot Generalized Linear Model for Switching Brain State Discovery}

\author{Chengrui Li \\
Georgia Institute of Technology \\
\texttt{cnlichengrui@gatech.edu} \\
\And
Soon Ho Kim \\
Georgia Institute of Technology \\
\texttt{soonhokim@gatech.edu} \\
\And
Chris Rodgers \\
Emory University \\
\texttt{christopher.rodgers@emory.edu} \\
\And
Hannah Choi \\
Georgia Institute of Technology \\
\texttt{hannahch@gatech.edu} \\
\And
Anqi Wu \\
Georgia Institute of Technology \\
\texttt{anqiwu@gatech.edu} \\
}

\iclrfinalcopy 
\begin{document}

\maketitle

\begin{abstract}
    Exposing meaningful and interpretable neural interactions is critical to understanding neural circuits. Inferred neural interactions from neural signals primarily reflect functional interactions. In a long experiment, subject animals may experience different stages defined by the experiment, stimuli, or behavioral states, and hence functional interactions can change over time. To model dynamically changing functional interactions, prior work employs state-switching generalized linear models with hidden Markov models (i.e., HMM-GLMs). However, we argue they lack biological plausibility, as functional interactions are shaped and confined by the underlying anatomical connectome. Here, we propose a novel prior-informed state-switching GLM. We introduce both a Gaussian prior and a one-hot prior over the GLM in each state. The priors are learnable. We will show that the learned prior should capture the state-constant interaction, shedding light on the underlying anatomical connectome and revealing more likely physical neuron interactions. The state-dependent interaction modeled by each GLM offers traceability to capture functional variations across multiple brain states. Our methods effectively recover true interaction structures in simulated data, achieve the highest predictive likelihood with real neural datasets, and render interaction structures and hidden states more interpretable when applied to real neural data.
\end{abstract}

\section{Introduction}
Unveiling meaningful and interpretable neural interaction structures is vital for comprehending neural circuits. Extensive research has investigated these interactions using statistical and information-theoretic methods like cross-correlogram \citep{jia2022multi}, mutual information \citep{houghton2019calculating}, Granger causality \citep{granger1969investigating}, transfer entropy \citep{schreiber2000measuring}, and generalized linear methods \citep{linderman2016bayesian}. 

Typically, the inferred neural interaction from neural signals primarily reflects functional interaction subject to variations in neural activity. Direct observation or inference of the anatomical connectome, encompassing axons, dendrites, and synapses that establish neural communication, is usually not feasible. Moreover, functional interaction, unlike anatomical connectome, varies with behavioral states and on much faster time scales than anatomical connectome which remains relatively stable over a short period of time. Functional networks of neurons, therefore, reflect dynamic modes of computation shaped by task and sensory inputs. Existing experimental results provide evidence suggesting that many neural systems can exhibit diverse and state-changing firing patterns given different sensory, perceptual, and behavioral states \citep{sherman2001tonic, haider2007enhancement, anderson2000stimulus, sanchez2000cellular, escola2011hidden}. 

To capture such time-varying functional interactions in multi-state systems, prior studies explored state-switching generalized linear models (GLMs) with hidden Markov models (HMMs), referred to as HMM-GLMs \citep{escola2011hidden, nadagouda2021switched, zhou2021nonlinear, morariu2022state}. These models introduce a discrete hidden variable representing the state of each time point, with each state equipped with its own GLM to capture neural interactions. However, we argue that such methods are not biologically plausible enough to capture functional interaction in multi-state neural systems. 

In fact, an interaction between a pair of neurons inferred from neural signals can reflect not only functional interaction but also anatomical connectome or synaptic connectivity. There exists experimental evidence manifesting degrees of correlations between functional and anatomical networks \citep{gencc2016functional,siegle2021survey}. It is thus plausible to assume that functional interaction is dynamically modulated by brain states while also being shaped and confined by the underlying anatomical connectome.

Incorporating these more biologically plausible assumptions, we introduce the one-hot HMM-GLM, a novel approach for capturing time-varying functional interactions in multi-state neural systems using an HMM-GLM framework. Unlike previous HMM-GLM methods 
that assume complete independence among GLMs in different states, we introduce a learnable prior for all states, constraining the search space for the interaction weight of each GLM derived from neural activity. This approach reveals more anatomically informative functional interactions between neurons.

The next question is how to impose the prior over GLMs. We first provide a solution using a shared Gaussian prior over the interaction weight matrices of GLMs for all states, denoted as Gaussian HMM-GLM. However, this Gaussian prior is relatively naive and doesn't explicitly connect functional interactions to the anatomical connectome. Accordingly, we provide a second solution that decomposes each GLM's weight matrix into a connection matrix and a strength matrix, with the connection matrix modeled by a one-hot encoding mechanism. Our prior is then imposed solely on the connection, not the entire weight matrix. We argue that the regulated connection matrices, guided by the prior, shed light on the underlying anatomical connectome, revealing more likely physical interactions of neurons. Meanwhile, less restricted strength matrices offer traceability to capture functional variations across multiple brain states. Our experimental results demonstrate that, when compared to alternatives, one-hot HMM-GLM accurately recovers true interaction structures in simulated data and achieves the highest predictive likelihood on test spike trains from two real neural datasets. Moreover, the uncovered interaction structures and hidden states are more interpretable compared with alternatives in real neural datasets.

\section{Method}
\textbf{Classic GLM}: 
We denote a spike train data as $\bm{X}\in\Nd^{T\times N}$ recorded from $N$ neurons across $T$ time bins, $x_{t,n}$ as the number of spikes generated by the $n$-th neuron in the $t$-th time bin, and $\bm x_{t}\in\mathbb{R}^{N\times 1}$ as the vector of spikes for all neurons at time $t$. When provided with $\bm X$, a classic GLM, with pre-defined basis functions, predicts the firing rates of the $n$-th neuron at the time bin $t$ as
\begin{equation}\label{eq:glm}
    f_{t,n} = \sigma\rbr{b_n + \sum_{n'=1}^N w_{n\gets n'}\cdot \rbr{\sum_{k=1}^K x_{t-k,n'}\phi_k}},\quad  \mbox{with spike } x_{t,n} \sim \mathrm{Poisson}(f_{t,n}),
\end{equation}
where $\sigma(\cdot)$ is a non-linear function (e.g., Softplus); $b_n$ is the background intensity of the $n$-th neuron; $w_{n\gets n'}$ is the weight of the influence from the $n'$-th neuron to the $n$-th neuron whose matrix form is $\bm W\in\Rd^{N\times N}$; $\bm \phi\in\Rd_+^K$ is the basis function summarizing history spikes from $t-K$ to $t-1$. The GLM finds the optimal $\bm W$ by maximizing the Poisson log-likelihood of the observed spikes.

\begin{figure}[!t]
    \centering
    \includegraphics[width=\linewidth]{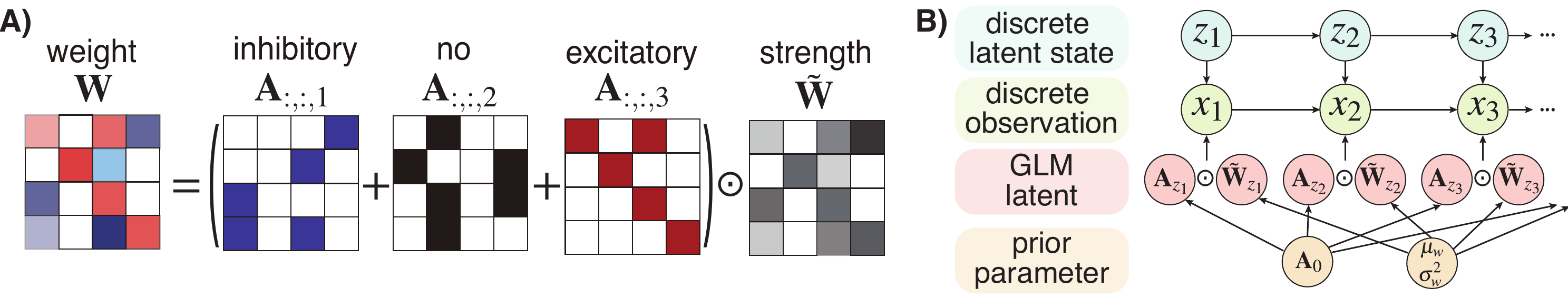}
    \caption{ A) A descriptive schematic of the weight matrix decomposition. B) The graphical model of the one-hot HMM-GLM.}
    \label{fig:diagram}
\end{figure}

\textbf{One-hot GLM}: 
We first introduce the novel one-hot GLM that produces a discrete connection matrix with type and a positive-valued strength matrix, i.e., 
\begin{equation}\label{eq:One-hot}
    {w}_{n\gets n'} = \sbr{(-1)a_{n\gets n',\inh} + (+1)a_{n\gets n',\exc}} \cdot \tilde{w}_{n\gets n'}.
\end{equation}
$\tilde{w}_{n\gets n'} \in\Rd_+$ is the strength of the weight. We define $\bm a_{n\gets n'} = [a_{n\gets n',\inh}, a_{n\gets n',\no},a_{n\gets n',\exc}]\in \Delta^2$ to be the type of the weight from neuron $n'$ to neuron $n$ corresponding to \{inhibitory, no connection, excitatory\}. $\bm a_{n\gets n'}$ is a soft one-hot encoding vector over a Simplex $\Delta^2 \coloneqq \{\bm a\in [0, 1]^3|\sum_{i=1}^3 a_i = 1\}$. The matrix and tensor forms are denoted as $\tilde{\bm W}\in\mathbb{R}_+^{N\times N}$ and $\bm A\in\{-1,0,1\}^{N\times N\times 3}$ respectively. Fig.~\ref{fig:diagram}A shows a schematic of the one-hot decomposition. 

\textbf{One-hot HMM-GLM}:
Next, we extend the one-hot GLM with an HMM (a schematic diagram in Fig.~\ref{fig:diagram}B). We assume there exist $S$ states underlying the functional interaction of neural activity. For each time $t$, we introduce a discrete latent variable $z_t\in\{1,\dots,S\}$, whose transition probability is $p(z_{t+1}|z_t) = \pi_{z_t,z_{t+1}}$ with a matrix form $\bm\varPi\in\Rd^{S\times S}$. Given a latent state $z_t$, we extend the notations for one-hot GLM in Eq.~\ref{eq:One-hot} to be $\bm W_{z_t}$, $\tilde{\bm W}_{z_t}$ and $\bm A_{z_t}$. Then the emission model is $p(x_{t,n}|z_t,\bm x_{1},\dots,\bm x_{t-1})=\mathrm{Poisson}(f_{t,n})$:
\begin{eqnarray}\label{eq:One-hot hmm-glm}
\quad f_{t,n} &=& \sigma\rbr{b_n + \sum_{n'=1}^N w_{z_t,n\gets n'}\cdot \rbr{\sum_{k=1}^K x_{t-k,n'}\phi_k}},\\
\mbox{and}\quad  {w}_{z_t,n\gets n'} &=& \sbr{(-1)a_{z_t,n\gets n',\inh} + (+1)a_{z_t,n\gets n',\exc}} \cdot \tilde{w}_{z_t,n\gets n'}.\nonumber
\end{eqnarray}
Note that the traditional HMM framework assumes that the emission probability distributions, similar to the transition probability distributions, are time-homogeneous, i.e., the emission model does not depend on any previous observations. Here we relax the assumption by introducing the dependence over the spike history, similar to the previous HMM-GLMs \citep{escola2011hidden}.

To impose the assumption that functional interactions across different states should share some common structure informing us about the underlying anatomical connectome, we impose a Gumbel-softmax prior over $\bm a_{s,n\gets n'}$, i.e., $\bm a_{s,n\gets n'}\sim \GSoftmax(\bm a_{0,n\gets n'}, \tau),\ \forall s\in\cbr{1,\dots, S}$, written out as
\begin{equation}\label{eq:gs}
    a_{s,n\gets n',\text{type}} = \frac{\exp\sbr{({\ln a_{0,n\gets n',\text{type}}+ g_{s,n\gets n',\text{type}}})/{\tau}}}{\sum_{\text{type}'\in\cbr{\inh,\no,\exc}} \exp\sbr{({\ln a_{0,n\gets n',\text{type}'} + g_{s,n\gets n',\text{type}'}})/{\tau}}},\quad \forall \text{ type}\in\cbr{\inh,\no,\exc}
\end{equation}
where $g_{s,n\gets n',\text{type}}\overset{\text{i.i.d.}}{\sim} \operatorname{Gumbel}(0,1)$. In practice, we can sample $g$ by sampling $u$ from $\operatorname{Uniform}(0,1)$ and computing $g = -\ln(-\ln(u))$. $\tau > 0$ is a temperature hyperparameter forcing $\bm a_{s,n\gets n'}$ to be a soft one-hot representation of the weight type. 
The tensor form of $\bm a_{0,n\gets n'}$ is denoted as $\bm A_0\in\mathbb{R}^{N\times N\times 3}$, which is a free-parameter matrix imposing the biological structure similarity over different states. Since $\bm A_0$ is a 3-way tensor with excitatory, inhibitory, and no connections, we consider it to well resemble synaptic connectivity. Consequently, if the synaptic connectivity is excitatory, its functional interaction is likely to be excitatory; and vice versa. The log density of the Gumbel-Softmax distribution is:
\begin{equation}
    \begin{split}
        \ln p(\bm a_{s,n\gets n'}|\bm a_{0,n\gets n'}) = & \Bigg[\ln 2 + 2\tau - 2 \ln\rbr{\sum_{\text{type}\in\cbr{\inh,\no,\exc}} \frac{a_{0,n\gets n',\text{type}}}{(a_{s,n\gets n',\text{type}})^\tau}} \\
        & + \sum_{\text{type}\in\cbr{\inh,\no,\exc}} \rbr{\ln a_{0,n\gets n',\text{type}} - (\tau+1)\ln (a_{s,n\gets n',\text{type}})}\Bigg].
    \end{split}
\end{equation}
Please refer to \cite{jang2016categorical} and \cite{maddison2016concrete} for a more detailed derivation. 

By introducing a Gumbel-Softmax prior over the connection matrix $\bm A$, we turn the parameter $\bm A$ into a latent variable. We also assume the strength $\tilde{\bm W}$ and the background intensity $b_n$ are random variables from some prior distributions. We put a Gaussian prior over the log of $\tilde{\bm W}$ to ensure its non-negativity and a Gaussian prior over $b_n$. The final generative model of one-hot HMM-GLM is
\begin{eqnarray}\label{eq:hmm-glm-gen}
z_{t+1}|z_t&\!\!\!\!\sim \!\!\!\!&\operatorname{Categorical}(\pi_{z_t,1},\dots,\pi_{z_t,S}),\quad \forall t\in\{1,\dots,T\}\nonumber\\
\bm a_{s,n\gets n'}&\!\!\!\!\sim \!\!\!\! &\GSoftmax(\bm a_{0,n\gets n'}, \tau), \quad\forall s\in\cbr{1,\dots, S},\ \forall n, n'\in\cbr{1,\dots, N}\nonumber\\
\ln \tilde{w}_{s,n\gets n'}&\!\!\!\!\sim \!\!\!\! &\Nc(\mu_w,\sigma^2_w), \quad\forall s\in\cbr{1,\dots, S},\ \forall n, n'\in\cbr{1,\dots, N}\\
b_n&\!\!\!\!\sim \!\!\!\! &\Nc(\mu_b,\sigma^2_b), \quad\forall n\in\cbr{1,\dots, N}\nonumber \\
x_{t,n}& \!\!\!\!\sim \!\!\!\! & \mbox{Poisson}(f_{t,n}(\bm x_1, \dots, \bm x_{t-1}, \bm A_{z_t},\tilde{\bm W}_{z_t},b_n)), \quad\forall t\in\cbr{1,\dots, T}, \forall n, n'\in\cbr{1,\dots, N}.\nonumber
\end{eqnarray}

\textbf{Gaussian HMM-GLM}: We can achieve another variant of HMM-GLM by using the weight $w_{s,n\gets n'}$ without decomposition and imposing a Gaussian prior $\Nc(w_{0,n\gets n'},\sigma^2)$ on the weight $w_{s,n\gets n'}$ with hyperparameter $\sigma^2$, $\forall s\in\cbr{1,\dots,S}$, referred to as Gaussian HMM-GLM. It is similar to one-hot HMM-GLM in the sense that they both assume that the state-dependent weights $\bm W_s$ share some common information ($\bm A_0$ for one-hot HMM-GLM and $\bm W_0$ for Gaussian HMM-GLM). The main difference is that Gaussian HMM-GLM does not differentiate the connection from the interaction strength. Therefore, the shared $\bm W_0$ incorporates both, while in one-hot HMM-GLM, thanks to the decomposition, $\bm A_0$ only imposes similarity over the connection, not the strength. The regulated connection matrices with their prior should inform us about the underlying anatomical connectome. The less restricted strength matrices provide us with sufficient traceability to capture functional variations across multiple brain states. We will show, in the experimental evaluation section, that a biologically plausible constraint like $\bm A_0$ in one-hot HMM-GLM is critical to obtaining meaningful inference and learning results.

\section{Inference}
Our generative model has four latent variables $\{z_t, \bm A_s, \ln  \tilde{\bm W}_s,b_n\}$. It requires a complex fully Bayesian inference approach to infer all the latent variables, which is usually very time-consuming and highly computationally intensive. We provide a Baum-Welch algorithm to solve the inference problem. In our Baum-Welch, we derive the posterior of $z_t$ in the E-step, and do maximum a posteriori estimation for all other latent variables given the estimated posterior distribution of $z_t$ in the M-step, i.e., we jointly optimize model parameters and latent variables in the M-step. The rationale is that the calculation of the posterior for $z_t$ is straightforward via forward-backward message passing, while the calculation of the posterior for $\bm A_s$ is very challenging and has no closed-form expression. We can certainly resort to a variational distribution to approximate the posterior for $\bm A_s$. However, since the prior of $\bm A_s$ is a Gumbel-Softmax distribution, it is unclear what parametric density function we should choose to serve as the approximated posterior distribution. Given these challenges, we only do the E-step for $z_t$ with forward-backward message passing. In the M-step, we optimize the model parameters $\{\bm\varPi,\bm A_0\}$ with $\{\bm A_s, \ln \mathbf{\tilde{W}}_s,b_n\}$, denoted as $\theta$ altogether. The hyperparemeter set is $\zeta=\{\mu_w,\sigma^2_w,\mu_b,\sigma^2_b,\tau\}$, which is pre-defined, detailed later. We also pre-define the basis function $\phi\in R_+^K$.

First, we infer the hidden state given $\theta^{\text{old}}$ with the forward-backward algorithm (E-step). In this step, we will omit $\theta^{\text{old}}$ for simplicity. We define $
\gamma_{z_t}(t) \coloneqq p(z_t|\bm X;\theta^{\text{old}}),\quad
        \xi_{z_{t-1},z_t}(t) \coloneqq p(z_{t-1},z_t|\bm X;\theta^{\text{old}})$, 
and define $
        \alpha_{z_t}(t) \coloneqq p(\bm x_1,\dots,\bm x_t,z_t),\quad
        \beta_{z_t}(t) \coloneqq p(z_{t+1},\dots,z_T|\bm x_1,\dots,\bm x_t,z_t)$. 
Then, we can obtain the relationship $
        \gamma_{z_t}(t) = \frac{\alpha_{z_t}(t)\beta_{z_t}(t)}{p(\bm X)} ,\quad
        \xi_{z_{t-1},z_t}(t) = \frac{\beta_{z_t}(t)p(\bm x_t|\bm x_1,\dots,\bm x_{t-1},z_t)\alpha_{z_{t-1}}(t-1)p(z_t|z_{t-1})}{p(\bm X)}$. 
$\alpha_{z_t}(t)$ and $\beta_{z_t}(t)$ can be computed iteratively as
\begin{equation}
    \vspace{-0.05in}
    \begin{cases}
        \alpha_{z_t}(t) = p(\bm x_t|\bm x_1,\dots,\bm x_{t-1},z_t) \sum_{z_{t-1}=1}^S \alpha_{z_{t-1}}(t) p(z_t|z_{t-1}), & \alpha_{z_1}(1) = p(z_1)p(\bm x_1|z_1) \\
        \beta_{z_t}(t) = \sum_{z_{t+1}=1}^S \beta_{z_{t+1}}(t+1)p(\bm x_{t+1}|\bm x_1,\dots,\bm x_t,z_{t+1})p(z_{t+1}|z_t), & \beta_{z_T}(T) = 1\nonumber
    \end{cases}
\end{equation}
resulting in $p(\bm X) = \sum_{z_T=1}^S \alpha_{z_T}(T)$. With this inferred posterior for $\bm z$, we can update $\theta$ in the M-step by maximizing
\begin{equation}
    \begin{split}
        Q(\theta,\theta^{\text{old}}) = & \Ed_{p(\bm z|\bm X;\theta^{\text{old}})} \ln p(\bm X,\bm z;\theta) = \sum_{\bm z} p(\bm z|\bm X;\theta^{\text{old}}) \ln p(\bm X,\bm z;\theta) \\
        = & \sum_{z_{1}=1}^S \gamma_{z_1}(1)\ln p(z_1;\theta) + \sum_{t=2}^T\sum_{z_{t-1}=1}^S\sum_{z_t=1}^S \xi_{z_{t-1},z_t}(t) \ln p(z_t|z_{t-1};\theta) \\
        & + \sum_{t=1}^T\sum_{z_t=1}^S \gamma_{z_t}(t) \ln p(\bm x_t|\bm x_1,\dots,\bm x_{t-1},z_t;\theta).
    \end{split}\nonumber
\end{equation}
More details about the inference can be found in Appendix A.

There are several key hyperparameters in $\zeta$ requiring pre-defining before inference. (1) Gumbel-Softmax temperature $\tau$: It is common to choose the temperature $\tau$ in Gumbel-softmax from $[0.1, 1]$. If $\tau$ is too large, the relaxation will be too soft; if $\tau$ is too small, numerical issues could arise. In our model, $\tau$ is used to force the soft one-hot close to one corner of the simplex, so we tried $\tau \in \{0.1, 0.2, 0.5\}$, and found that the result of the one-hot HMM-GLM is not sensitive to $\tau$ in this range. Given that the selection of $\tau$ is insensitive to different datasets, we fix $\tau=0.2$, which is a common moderate choice. (2) Generative hyperparameters $\{\mu_w,\sigma_w^2,\mu_b,\sigma_b^2\}$: we chose $\mu_w = -5, \sigma_w = 2$ and $\mu_b = 0, \sigma_b = 2$ since this set provides noninformative priors for the strength/weight and the background intensity in GLMs, and hence the inference is insensitive to different datasets.

\section{Experimental evaluation}
\paragraph{Models for comparison.} We will compare our methods and state-of-the-art baseline methods on one simulated data and two real neural datasets:\\
$\bullet\quad$ \textbf{GLM} \citep{pillow2008spatio}: The most original model for discovering neural interactions, without the multiple-state assumption. \\
$\bullet\quad$\textbf{HMM Corr} \citep{engel2016selective}: An HMM for discovering state switches from spike train data. Since this method cannot find neural connectivities but only the latent states, we use a correlation-based method, i.e., cross-correlogram (CCG) to find the connectivities in each inferred state. \\
$\bullet\quad$ \textbf{HMM Bern} \citep{ashwood2022mice}: Similar to the HMM Corr, but uses the Bernoulli rather than Poisson distribution to model the spike count in each time bin. \\
$\bullet\quad$ \textbf{HG} \citep{escola2011hidden}: The classic HMM-GLM (HG) model, which is the only existing model that both infers states and learns neural connectivities. \\
$\bullet\quad$ \textbf{GHG} (our method): We denote Gaussian HMM-GLM as GHG. \\
$\bullet\quad$ \textbf{OHG} (our method): We denote one-hot HMM-GLM as OHG. \\
$\bullet\quad$ \textbf{HG-L1} and \textbf{GHG-L1}: Given that the one-hot mechanism implicitly imposes sparsity on the weight matrix, concerns may arise regarding whether the imposition of sparsity solely accounts for OHG's superiority. To address this, we will conduct two comparisons: one by adding an L1 penalty to the weight of HG, denoted as HG-L1, and another to GHG, denoted as GHG-L1. We will determine the L1 penalty coefficient through validation.

\paragraph{Metrics.} We use the following metrics to report performances from different methods:\\
$\bullet\quad$ \textbf{LL}. The log-likelihood on the test set. A better model should have a stronger ability to predict future spiking events. Note that this is the only metric that can be used on real-world datasets, since there are no true states and neural connectivity available for real-world datasets. \\
$\bullet\quad$ \textbf{State accuracy}. The average accuracy of the inferred states across all time bins. This is only applicable to the simulated dataset where we know the true hidden states. \\
$\bullet\quad$ \textbf{Weight error}. The error of the learned weight matrices in all states. Note that there is no weight error for HMM Corr and HMM Bern. Since their learned weights are from CCG, the weights cannot be compared with weights in the GLM model. This is only applicable to the simulated dataset.\\
$\bullet\quad$ \textbf{Connection accuracy}. The balanced accuracy of the learned connection matrices in all states. For models without connection matrices explicitly modeled, we use
\begin{equation}\label{eq:bin}
    \bm a_{s,n\gets n'} = \begin{cases}
        \rbr{0, 1-\frac{w_{s,n\gets n'}}{\max_{s,n,n'}{w_{s,n\gets n}}},\frac{w_{s,n\gets n'}}{\max_{s,n,n'}{w_{s,n\gets n}}}}, & w_{s,n\gets n'} \geqslant 0 \\
        \rbr{\frac{w_{s,n\gets n'}}{\min_{s,n,n'}{w_{s,n\gets n}}}, 1-\frac{w_{s,n\gets n'}}{\min_{s,n,n'}{w_{s,n\gets n}}}, 0}, & w_{s,n\gets n'} < 0
    \end{cases}
\end{equation}
to obtain the connection matrix from the learned weight matrix. We choose Eq.~\ref{eq:bin} since it is an automatic way with a reasonable rationale. We can also use a pre-defined threshold to obtain the connection matrix, but the accuracy of the connection matrices is very sensitive to the thresholding technique (see Appendix \ref{appendix:threshold}). In real neural data analysis, when we don't have the ground-truth connection matrices, we cannot even use such an accuracy metric to select the optimal threshold value. This demonstrates that the explicit connection matrices from the one-hot HMM-GLM provide a succinct expression requiring no pre-defined thresholds but render satisfactory estimation. This is only applicable to the simulated dataset.\\
$\bullet\quad$ \textbf{Connection prior accuracy}. Except for one-hot HMM-GLM, the connection prior is obtained by first averaging the weight matrices across all states and then fitting the averaged weight to Eq.~\ref{eq:bin}. This is only applicable to the simulated dataset.


\subsection{Application to simulated data}
\paragraph{Dataset.} We first compare different models on a 5-state-20-neuron synthetic dataset with 10 independent trials. For each trial, we generate 20 spike sequences of length $T=5000$. Each spike sequence is generated from the generative model in Eq.~\ref{eq:hmm-glm-gen}, with $\pi_{s,s'} = 0.005 + 0.975\cdot\mathbbm{1}[s=s']$, $\tau=0$, $\mu_w = -5, \sigma_w^2=1.5$, and $\mu_b = 0, \sigma_b^2=0.0008$. We sample $\bm a_{0,n'\gets n}$ from $\operatorname{Dirichlet}(0.1, 0.8, 0.1),\ \forall n,n'\in \cbr{1,\dots, 20}$. Note that instead of using the Gumbel-Softmax to generate $\bm a_{s,n'\gets n}$, we sample it from a Categorical distribution, i.e., $\bm a_{s,n'\gets n} \sim \operatorname{Categorical}(\bm a_{0,n'\gets n},0),\ \forall s\in\cbr{1,\dots,5},\ \forall n,n'\in \cbr{1,\dots, 20}$. It actually introduces some mismatching generative procedures compared with Eq.~\ref{eq:hmm-glm-gen}. Note that when $\tau=0$, all $\bm A_s$ in this data generating model are hard one-hot encodings i.i.d. sampled from $\bm A_0$. For each trial, we train different models on the training set consisting of the first 10 sequences, and test on the test set consisting of the remaining 10 sequences.

\begin{table}[t]
    \centering
    \begin{tabular}{cccccc}
        \toprule
        method & LL $\uparrow$ & state acc $\uparrow$ & weight error $\downarrow$ & con acc $\uparrow$ & con prior acc $\uparrow$ \\
        \midrule
        GLM & -8.43\small{($\pm$0.18)} & nan\small{($\pm$nan)} & 24.71\small{($\pm$0.19)} & 43.12\small{($\pm$0.46)} & 44.81\small{($\pm$0.61)} \\
        HMM Corr & -22.53\small{($\pm$0.64)} & 42.84\small{($\pm$1.47)} & nan\small{($\pm$nan)} & 34.04\small{($\pm$0.12)} & 15.45\small{($\pm$2.49)} \\
        HMM Bern & -5.68\small{($\pm$0.23)} & 87.95\small{($\pm$0.93)} & nan\small{($\pm$nan)} & 36.25\small{($\pm$0.25)} & 40.70\small{($\pm$1.53)} \\
        HG & -5.49\small{($\pm$0.58)} & 37.73\small{($\pm$2.80)} & 109.67\small{($\pm$2.63)} & 34.17\small{($\pm$0.08)} & 40.91\small{($\pm$0.48)} \\
        HG-L1 & 9.14\small{($\pm$0.18)} & 91.60\small{($\pm$0.96)} & 23.14\small{($\pm$0.08)} & 37.47\small{($\pm$0.18)} & 48.44\small{($\pm$0.57)} \\
        GHG & 8.58\small{($\pm$0.19)} & 91.80\small{($\pm$0.92)} & 21.54\small{($\pm$0.15)} & 42.53\small{($\pm$0.22)} & 48.93\small{($\pm$0.54)} \\
        GHG-L1 & 9.77\small{($\pm$0.20)} & 92.08\small{($\pm$0.89)} & 14.16\small{($\pm$0.07)} & 41.08\small{($\pm$0.22)} & 46.98\small{($\pm$0.60)} \\
        OHG & \textbf{14.64}\small{($\pm$0.23)} & \textbf{92.75}\small{($\pm$0.87)} & \textbf{10.99}\small{($\pm$0.21)} & \textbf{73.90}\small{($\pm$0.52)} & \textbf{80.60}\small{($\pm$0.59)} \\
        \bottomrule
    \end{tabular}
    \caption{The quantitative results with 5 metrics on the synthetic dataset.}
    \label{tab:synthetic}
\end{table}

\begin{figure}[t]
    \centering
    \includegraphics[width=\textwidth]{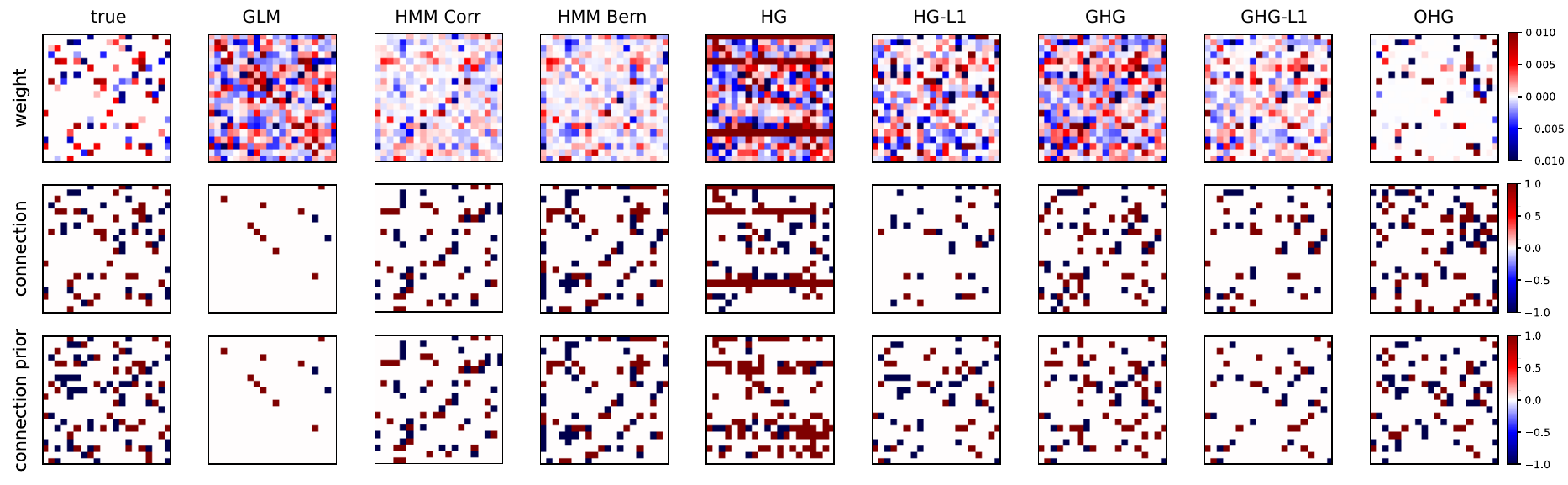}
    \vspace{-0.2in}
    \caption{Visualization of weight $\bm W_2$ (top row), connection $\bm A_2$ (middle row), and connection prior $\bm A_0$ (bottom row) for all methods corresponding to state 2 ($S=5$ in total) learned from one trial of the synthetic dataset.}
    \label{fig:synthetic_weights}
\end{figure}

We show the quantitative results in Tab.~\ref{tab:synthetic} and the learned neural connectivities in Fig.~\ref{fig:synthetic_weights}. 
From Tab.~\ref{tab:synthetic}, we can tell that our OHG is the best in terms of all five metrics. Next, we make use of the neural connectivities learned by different models (Fig.~\ref{fig:synthetic_weights}) to analyze the results. Since there are $S=5$ different states, one-state GLM is only able to capture an ``average'' estimation among the $5$ states. For HMM Corr and HMM Bern, the learning procedure is decoupled into two steps, inferring hidden states and estimating the neural connectivities on each inferred state. Although the inferred hidden state from HMM Bern is acceptable, the estimated connection matrix in each state and the connection prior are still bad. For HG, the poor performance is mainly from an incorrect estimation of the transition matrix, which leads to a bad inference of the hidden state sequence (Fig.~\ref{fig:synthetic_states_appendix} in Appendix~\ref{fig:synthetic_states_appendix}) and hence results in a wrong weight and connection estimation. Comparing HG with GHG and OHG, we conclude that a constraint (i.e., the connection prior) on different states is necessary to get a stable result. The shared information between different states can help prevent the inferred states and the weights in different states from falling into extremes or bad local optima. Adding an L1 penalty could suppress some of the noisy weights but is still not helpful for estimating connections in each state and the shared connection prior, as L1 does not enhance discrimination between weak and no connections. The main difference between GHG and OHG is their weight and connection estimation. We can tell that GHG still has many noisy non-zero weights. With the one-hot setting in OHG, the sparsity of the network is easily learned, and connections with zero interactions are successfully suppressed, which leads to a lower weight error and better connection accuracy (the weights, connections, and the connection prior learned by OHG match the true the best in Fig.~\ref{fig:synthetic_weights}).

\subsection{Applications to electrophysiology data}
\subsubsection{Prefrontal cortex during a contingency task}
We first apply different models to a prefrontal cortex (PFC) dataset \citep{peyrache2018activity, peyrache2009replay}\footnote{\url{https://crcns.org/data-sets/pfc/pfc-6}}. Neural spike trains were collected while a rat learned a behavioral contingency task. During recording, the animal performed a trial for about 4 secs and then took a short break for about 24 secs. The spike train data used for learning and testing is segmented from the long session. Each sequence starts from 5 seconds before a behavior starts and lasts for 10 seconds after the start. Hence, each sequence corresponds to a behavioral trial. We use $\frac{2}{3}$ of the neural sequences as the training set and the remaining $\frac{1}{3}$ as the test. The neural spikes are binned into 750 time bins with bin size = 20 ms. Since we do not know the true number of hidden states, we try $S\in \cbr{2,3,4,5}$.

\begin{figure}[t]
    \centering
    \includegraphics[width=0.9\textwidth]{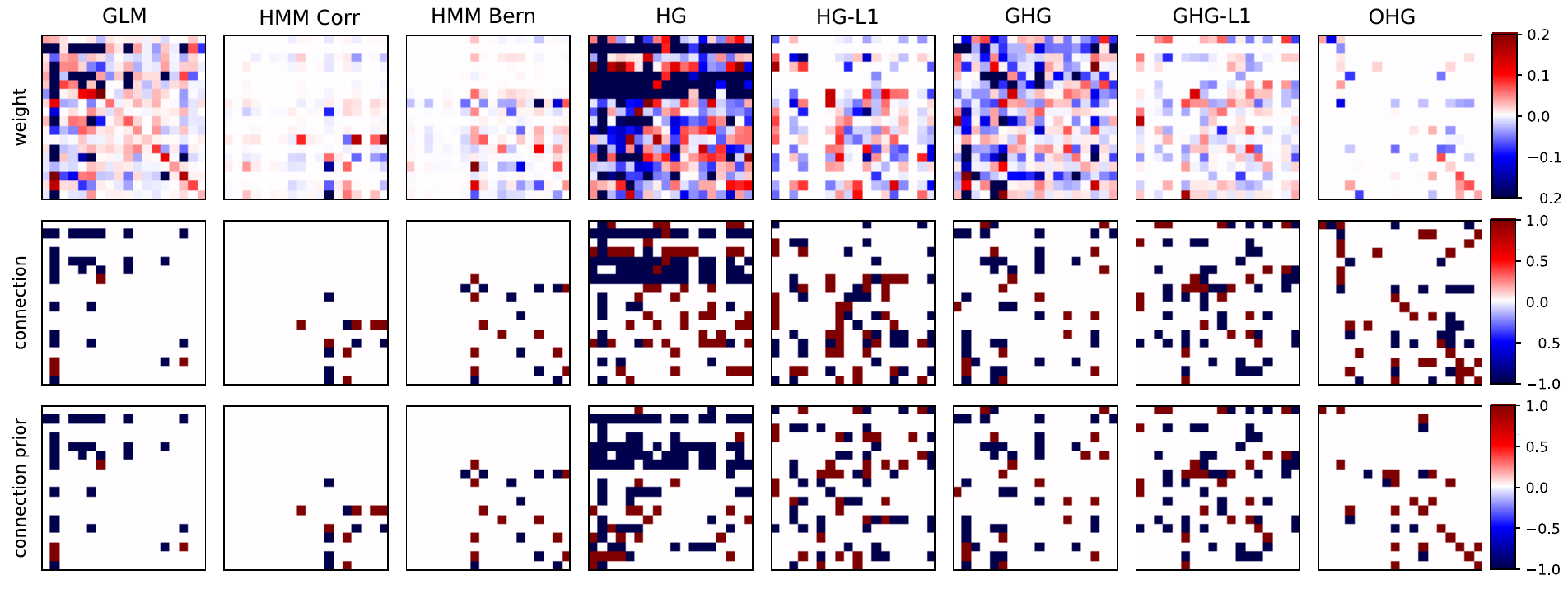}
    \vspace{-0.1in}
    \caption{Visualization of weight $\bm W_4$ (top row), connection $\bm A_4$ (middle row), and connection prior $\bm A_0$ (bottom row) for all methods corresponding to the state 1 ($S=4$ in total) learned from the PFC-6 dataset.}
    \label{fig:pfc6_weights}
\end{figure}

Tab.~\ref{tab:pfc6} shows that the test log-likelihoods of OHG with all different numbers of states are consistently better than others. Fig.~\ref{fig:pfc6_weights} shows an example of the weights and connections estimated by different models. 
For HG, the learned weight matrices are pretty dense and noisy, resulting in a bad log-likelihood on the test set. For GHG, the weight is less dense but still noisy. Adding L1 penalties to HG and GHG is helpful for reducing some noisy weight entries, but still not helpful for discriminating between weak connection and no connection. Using OHG, we can get a much clearer strength-connection decomposition and also obtain a connection prior. The global restriction provided by the connection prior shapes the functional interactions as the anatomical connectome does, which improves the log-likelihood of the model on the test set. Note that GLM actually achieves a reasonably good result, only worse than OHG. It indicates that in such real-world scenarios, functional interactions in different states indeed share a global static connection prior (may reflect the anatomical connectome), outweighing the functional differences between different states and hence should be taken into account.

\begin{figure}[t]
    \centering
      \includegraphics[width=\textwidth]{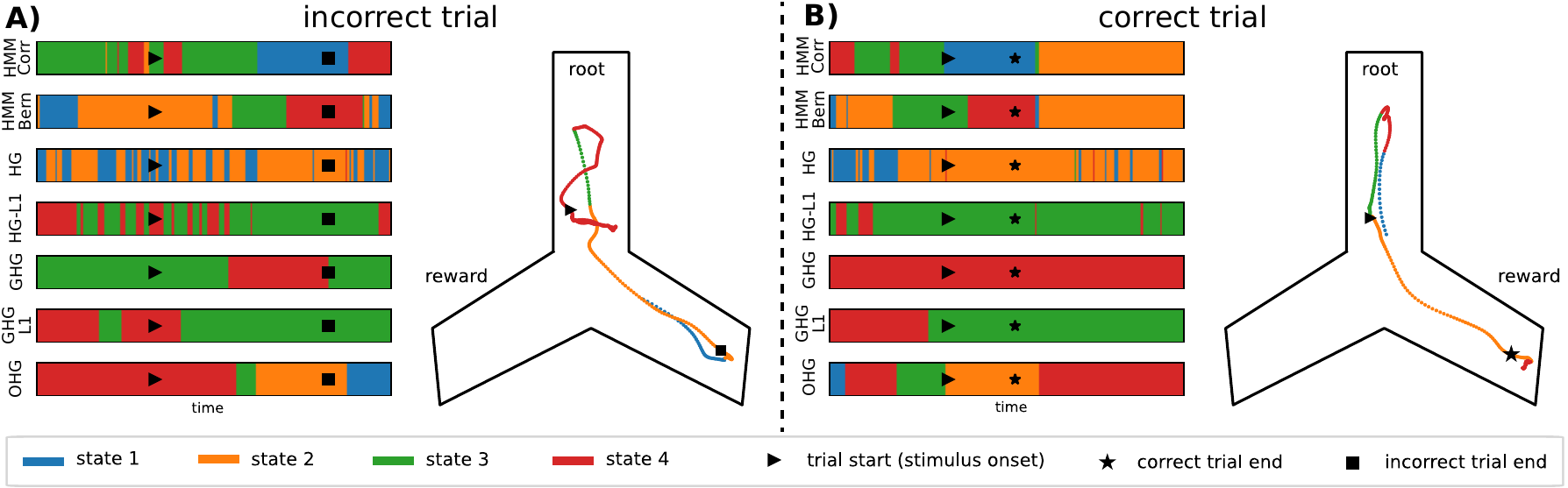}
   \vspace{-0.2in} 
   \caption{Inferred hidden states of an incorrect trial (A) and a correct trial (B) from various models including HG, GHG, and OHG. The rat trajectory on the right-hand side of each one is colored according to the hidden states inferred from OHG.}
    \label{fig:pfc6_states}
\end{figure}

Although there is no ground truth of hidden states, we can integrate the behavioral data to analyze the inferred hidden states from different models. Pick 4 states as an example. In Fig.~\ref{fig:pfc6_states}, we plot the hidden state prediction of one incorrect trial (Fig.~\ref{fig:pfc6_states}(A)) and one correct trial (Fig.~\ref{fig:pfc6_states}(B)). We also plot the corresponding rat movement on the right-hand side. As previously observed, HG continues to yield a state prediction characterized by significant noise and limited interpretability. Although the number of hidden states is set as $S=4$, GHG only infers two effective hidden states. The transition from state 4 to state 3 typically happens when the rat turns back at the wrong target location. However, OHG is able to find four explainable effective hidden states. Before each trial, the rat goes back to the root of the Y-shaped maze (starting point), corresponding to state 4. Then the rat turns around at the starting point and goes forward to the turning point of the Y-shaped maze, corresponding to state 3. After making the decision, the rat enters into state 2 in one arm of the Y-shaped maze, to reach the destination. If the rat goes to the correct target location, it gets a reward at the target and the rat will stay in state 4 for a long while. But if the rat goes to the incorrect target location, there is no reward and the rat will go back immediately, corresponding to state 1. The state explanation of the OHG is reflected in the colored rat trajectory in Fig.~\ref{fig:pfc6_states} (the trajectory is colored by the state predicted by OHG). Note that the state patterns for correct and incorrect trials are not from cherry-picking. We do observe similar state transitions among other more correct and incorrect trials, which can be checked and validated in Fig.~\ref{fig:pfc6_states_appendix} in Appendix \ref{appendix:pfc6}.

\begin{table}[t]
    \centering
    \begin{tabular}{ccccc}
        \toprule
        method & 2 states & 3 states & 4 states & 5 states \\
        \midrule
        HMM Corr & -37.11\small{($\pm$0.00)} & -36.60\small{($\pm$0.00)} & -36.53\small{($\pm$0.00)} & -36.68\small{($\pm$0.00)} \\
        HMM Bern & -36.89\small{($\pm$0.00)} & -36.57\small{($\pm$0.00)} & -36.38\small{($\pm$0.00)} & -36.38\small{($\pm$0.00)} \\
        HG & -37.30\small{($\pm$0.05)} & -37.61\small{($\pm$0.17)} & -37.22\small{($\pm$0.14)} & -36.98\small{($\pm$0.19)} \\
        HG-L1 & -36.91\small{($\pm$0.01)} & -36.90\small{($\pm$0.02)} & -36.73\small{($\pm$0.09)} & -36.63\small{($\pm$0.13)} \\
        GHG & -37.17\small{($\pm$0.00)} & -37.11\small{($\pm$0.01)} & -37.12\small{($\pm$0.00)} & -37.11\small{($\pm$0.00)} \\
        GHG-L1 & -36.94\small{($\pm$0.00)} & -36.88\small{($\pm$0.00)} & -36.83\small{($\pm$0.00)} & -36.77\small{($\pm$0.00)} \\
        OHG & \textbf{-35.92}\small{($\pm$0.02)} & \textbf{-35.79}\small{($\pm$0.02)} & \textbf{-35.77}\small{($\pm$0.03)} & \textbf{-35.71}\small{($\pm$0.03)} \\
        \bottomrule
    \end{tabular}
    \caption{The log-likelihood on the test set for different methods and different numbers of states of the PFC-6 dataset. The result from the one-state GLM is -36.35\small{($\pm$0.00)}.}
    \label{tab:pfc6}
\end{table}

\vspace{-0.1in}
\subsubsection{Barrel cortex during whisking}
\begin{figure}[!t]
    \centering
    \includegraphics[width=1\textwidth]{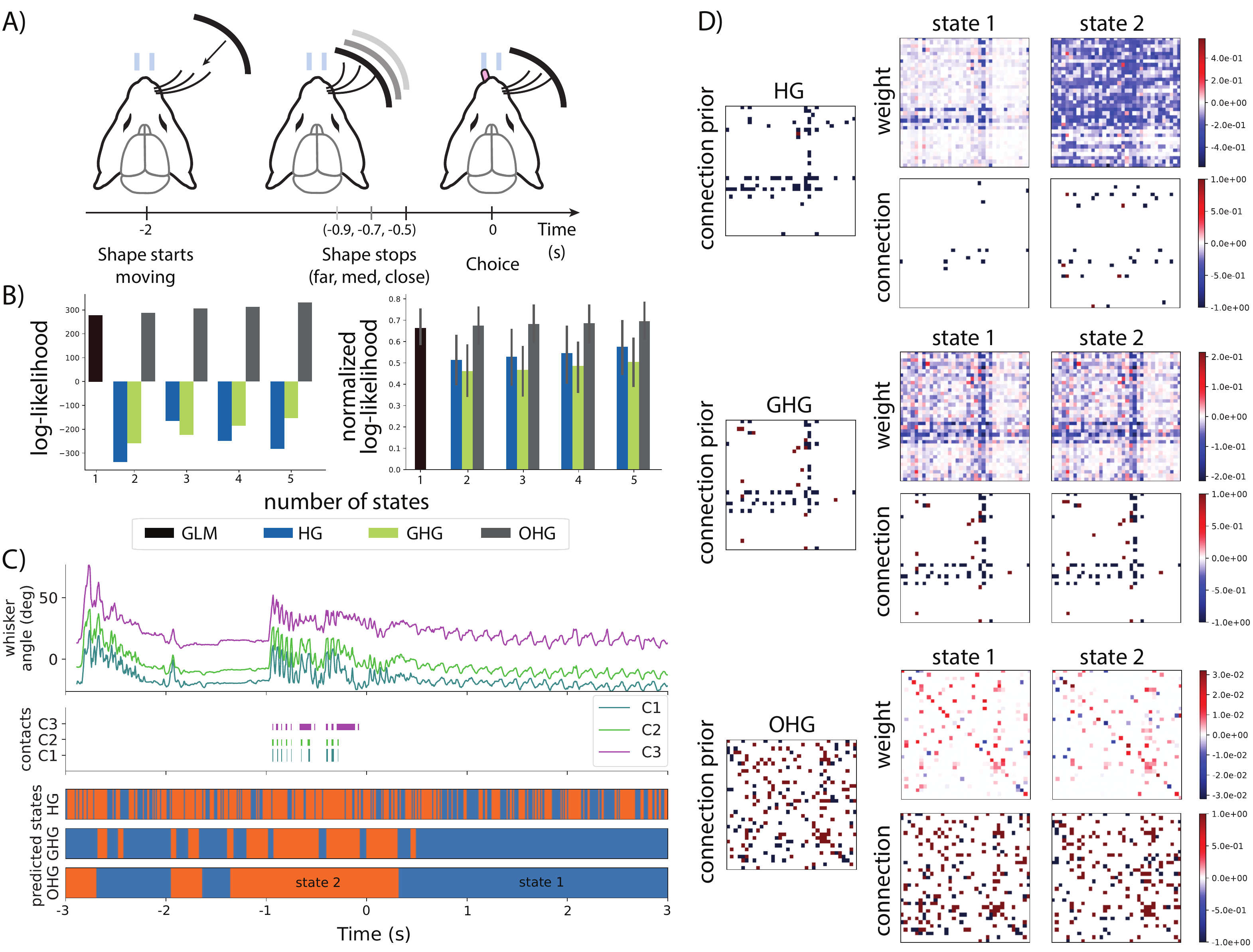}
    \vspace{-0.3in}
    \caption{A) Experimental setup for the whisking task (adapted from \cite{nogueira2023geometry}). B) Test (normalized) log-likelihoods given different methods. Error bars are not shown for raw log-likelihood (left) due to extremely high session-by-session variation. To account for this, the normalized log-likelihood is also shown (right). C) Example trial from a discrimination task. Whisker positions (top panel), whisker contacts with the object (middle), and the probability of the state being state 1 (bottom). Here $t=0$~s is the time at which the response window is opened (after which the lick direction of the mouse is considered as its decision), and the stimulus is presented at approximately $t=-1$~s. D) Weights, connections, and connection priors of the three models in the example session shown in C.}
    \label{fig:chris}
\end{figure}

\vspace{-0.1in}
\paragraph{Dataset.} We next apply the methods to electrode recordings of the somatosensory (barrel) cortex in mice during a shape discrimination task~\citep{rodgers2021sensorimotor, rodgers2022detailed, nogueira2023geometry} (Fig.~\ref{fig:chris}A). Mice were trained to discriminate concave from convex shapes using only their whiskers. In particular, the mice are required to actively whisk in order to make contact with the object; a high-speed video of whisker motion was collected, allowing analysis of the active movement of the whiskers to sense the environment. Here we use 27 sessions from 5 different mice. The number of recorded neurons varies from 10 to 44 across sessions. Six seconds from each trial is included in the analysis, and spike trains are discretized with a time bin of 3~ms. The first 30 trials are used in the analysis of each session of which 10 randomly selected trials form the test set when evaluating the test log-likelihood, and the remaining 20 trials are used for training the model.


Given that we do not have good knowledge about the behavioral states, we try different numbers of hidden states for the barrel cortex data, i.e., $S=\{2,3,4,5\}$. The log-likelihoods of the models fit to the barrel cortex dataset show similar trends to the PFC dataset; OHG consistently has the highest log-likelihood, and GHG generally exhibits greater log-likelihood compared to the base model across different numbers of hidden states (See Fig.~\ref{fig:chris}B).

Fig.~\ref{fig:chris}C shows whisker positions, contacts, and predicted hidden state transitions of each model. We select the case of $S=2$ hidden states here for visualization. While the log-likelihood of OHG increases as $S$ increases to 5, for $S>2$, there are many sessions with rarely occupied states, and the distinction between states becomes subtle. Results for 3-5 states are shown in Appendix \ref{appendix:barrel}. When two states are assumed, it is typically observed that one of the states inferred by GHG and OHG coincides with active whisking events during which contacts occurred, while the states predicted by the naive model switch very frequently. 

While GHG and OHG correlated with whisking events similarly, the durations of the predicted states are different (Fig.~\ref{fig:chris}). OHG predicts stable states with duration over 1~s that persist over whisking cycles, while the inferred states of GHG switch rapidly with short duration ($<0.1$ s). The OHG thus better captures sustained whisking cycles (\cite{deschenes2012sniffing, rodgers2021sensorimotor}). Fig.~\ref{fig:chris}D further shows the weights and connection matrices estimated by each model for the same session shown in Fig.~\ref{fig:chris}C. As in the PFC dataset, we observe that only OHG learns sparse weight matrices, while the ones learned by GHG and HG are denser and noisier.

We further test the idea that the states predicted by OHG and GHG are related to the active whisking events.
We compute the frequency with which whisker contacts are initiated in each state, and perform a chi-squared test against the expected frequencies if no relation between the states and contacts is assumed.
Among 11 sessions where all three models result in predicted state frequencies that are not completely skewed (the least frequent state was predicted in at least $5\%$ of the time steps), the null hypothesis is rejected ($p<0.001$) in 6 sessions (54\%) for HG and in 8 sessions (73\%) for both GHG and OHG. 
Furthermore, across all sessions, we compute the sum of all elements in the weight matrix $\bm W$ of the state associated with whisker contacts and that of the other state.
When comparing the distribution of total weight between whisking and non-whisking states, OHG results in a significant increase of the weights during whisking states ($p=0.008$, two-sided Wilcoxon rank-sum test), while GHG and HG do not ($p>0.1$). 
This suggests that OHG is capable of detecting shifts in functional interaction tied to switching behavioral states.

\section{Conclusion}
We develop a novel one-hot HMM-GLM (OHG) to estimate time-varying functional interaction in multi-state neural systems. The newly proposed OHG decomposes the traditional weight matrix in GLMs into a discrete connection matrix with type and a positive-valued strength matrix. Such a decomposition is critical when applied to state-switching neural interaction discovery. When building OHG, we place a common Gumbel-Softmax prior over the connection matrix for each state, enforcing the connection matrices to learn shared information. We argue that the regulated connection matrices with their shared prior should inform us about underlying anatomical connectome and thus uncover the ``more likely'' physical interactions between neurons. For the strength matrix, we allow it to change freely without a shared prior across states. The less restricted strength matrices will provide us with sufficient traceability to capture functional variations across multiple brain states. We argue that OHG is more biologically plausible given the aforementioned benefits. We show in the experiment that when compared with alternatives, OHG infers better connectivity and hidden states. It not only accurately recovers the true connectivity for simulated data but also achieves the best predictive likelihood on test spike trains for a PFC dataset and a barrel cortex dataset. The uncovered connectivity and hidden state sequence with OHG are more interpretable for these real neural datasets. 


\subsubsection*{Acknowledgments}
This work was supported by a Seed Grant: Forming Teams from Georgia Institute of Technology, and the National Eye Institute of the National Institutes of Health under Award Number R00 EY030840 and a Sloan Research Fellowship to H.C. The content is solely the responsibility of the authors, and does not necessarily represent the official views of the National Institutes of Health.

\bibliography{ref}
\bibliographystyle{iclr2024_conference}

\appendix
\section{Appendix}
\subsection{Inference and learning algorithms for HMM-GLM}
\subsubsection{Forward-backward inference}
In this part, we compute the posterior probability given the old parameter $\theta^{\text{old}}$, which is the E-step of the EM algorithm. Define
\begin{equation}
    \begin{cases}
        \gamma_{z_t}(t) \coloneqq p(z_t|\bm X;\theta^{\text{old}}) \\
        \xi_{z_{t-1},z_t}(t) \coloneqq p(z_{t-1},z_t|\bm X;\theta^{\text{old}})
    \end{cases}
\end{equation}
where $z_t$ indexes one of the $S$ different states.

Define
\begin{equation}
    \begin{cases}
        \alpha_{z_t}(t) \coloneqq p(\bm x_1,\dots,\bm x_t,z_t) \\
        \beta_{z_t}(t) \coloneqq p(z_{t+1},\dots,z_T|\bm x_1,\dots,\bm x_t,z_t)
    \end{cases}
\end{equation}
and we have
\begin{equation}\label{eq:gamma-alpha,beta}
    \begin{split}
        \gamma_{z_t}(t) = & \frac{p(\bm X,z_t)}{p(\bm X)} \\
        = & \frac{p(\bm x_1,\dots,\bm x_t,z_t)p(\bm x_{t+1},\dots,\bm x_T|\bm x_1,\dots,\bm x_t,z_t)}{p(\bm X)} \\
        = & \frac{\alpha_{z_t}(t)\beta_{z_t}(t)}{p(\bm X)}
    \end{split}
\end{equation}
\begin{equation}
    \begin{split}
        \alpha_{z_t}(t) = & p(\bm x_1,\dots,\bm x_t,z_t) \\
        = & p(\bm x_t|\bm x_1,\dots,\bm x_{t-1},z_t) p(\bm x_1,\dots,\bm x_{t-1},z_t) \\
        = & p(\bm x_t|\bm x_1,\dots,\bm x_{t-1},z_t) \sum_{z_{t-1}} p(\bm x_1,\dots,\bm x_{t-1},z_{t-1},z_t) \\
        = & p(\bm x_t|\bm x_1,\dots,\bm x_{t-1},z_t) \sum_{z_{t-1}} p(\bm x_1,\dots,\bm x_{t-1},z_t|z_{t-1})p(z_{t-1}) \\
        = & p(\bm x_t|\bm x_1,\dots,\bm x_{t-1},z_t) \sum_{z_{t-1}} p(\bm x_1,\dots,\bm x_{t-1}|z_t,z_{t-1})p(z_t|z_{t-1})p(z_{t-1}) \\
        = & p(\bm x_t|\bm x_1,\dots,\bm x_{t-1},z_t) \sum_{z_{t-1}} p(\bm x_1,\dots,\bm x_{t-1}|z_{t-1})p(z_{t-1})p(z_t|z_{t-1}) \\
        = & p(\bm x_t|\bm x_1,\dots,\bm x_{t-1},z_t) \sum_{z_{t-1}} \alpha_{z_{t-1}}(t)p(z_t|z_{t-1})
    \end{split}
\end{equation}
with initial condition
\begin{equation}
    \alpha_{z_1}(1) = p(\bm x_1,z_1) = p(z_1)p(\bm x_1|z_1)
\end{equation}
\begin{equation}
    \begin{split}
        \beta_{z_t}(t) = & p(\bm x_{t+1},\dots,\bm x_T|\bm x_1,\dots,\bm x_t,z_t) \\
        = & \sum_{z_{t+1}} p(\bm x_{t+1},\dots,\bm x_T,z_{t+1}|\bm x_1,\dots,\bm x_t,z_t) \\
        = & \sum_{z_{t+1}} p(\bm x_{t+1},\dots,\bm x_T|\bm x_1,\dots,\bm x_t,z_{t+1},z_t) p(\bm z_{t+1}|\bm x_1,\dots,\bm x_t,z_t) \\
        = & \sum_{z_{t+1}} p(\bm x_{t+1},\dots,\bm x_T|\bm x_1,\dots,\bm x_t,\bm z_{t+1}) p(\bm z_{t+1}|z_t) \\
        = & \sum_{z_{t+1}} p(\bm x_{t+2},\dots,\bm x_T|\bm x_1,\dots,\bm x_t,\bm x_{t+1},z_{t+1}) p(\bm x_{t+1}|\bm x_1,\dots,\bm x_t,z_{t+1}) p(z_{t+1}|z_t) \\
        = & \sum_{z_{t+1}} \beta_{z_{t+1}}(t+1) p(\bm x_{t+1}|\bm x_1,\dots,\bm x_t,z_{t+1}) p(z_{t+1}|z_t)
    \end{split}
\end{equation}
with initial condition $\beta_{z_T}(T) = 1$ since in Eq.~\ref{eq:gamma-alpha,beta}
\begin{equation}
    \gamma_{z_T}(T) = p(z_T|\bm X) = \frac{p(\bm X,z_T)\beta_{z+T}(T)}{p(\bm X)} \equiv \frac{p(\bm X,z_T)}{p(\bm X)}
\end{equation}

If we sum both sides of Eq.~\ref{eq:gamma-alpha,beta}
\begin{equation}
    1 = \sum_{z_t}\gamma_{z_t}(t) = \frac{\sum_{z_t}\alpha_{z_t}(t)\beta_{z_t}(t)}{p(\bm X)} \implies p(\bm X) = \sum_{z_t}\alpha_{z_t}(t)\beta_{z_t}(t)
\end{equation}
and we can simply use $p(\bm X) = \sum_{z_T}\alpha_{z_T}(T)$ when $t = T$.

\begin{equation}
    \begin{split}
        \xi_{z_{t-1},z_t}(t) = & p(z_{t-1},z_t|\bm X) \\
        = & \frac{p(\bm X|z_{t-1},z_t)p(z_{t-1},z_t)}{p(\bm X)} \\
        = & \frac{p(\bm X|z_{t-1},z_t) p(z_t|z_{t-1}) p(z_{t-1})}{p(\bm X)} \\
        = & \frac{p(\bm x_t,\dots,\bm x_T|\bm x_1,\dots,\bm x_{t-1},z_{t-1},z_t) p(\bm x_1,\dots,\bm x_{t-1}|z_{t-1},z_t) p(z_t|z_{t-1}) p(z_{t-1})}{p(\bm X)} \\
        = & \frac{p(\bm x_{t+1},\dots,\bm x_T|\bm x_1,\dots,\bm x_t,z_{t-1},z_t) p(\bm x_t|\bm x_1,\dots,\bm x_{t-1},z_{t-1},z_t) \alpha_{z_{t-1}}(t-1) p(z_t|z_{t-1})}{p(\bm X)} \\
        = & \frac{p(\bm x_{t+1},\dots,\bm x_T|\bm x_1,\dots,\bm x_t,z_t) p(\bm x_t|\bm x_1,\dots,\bm x_{t-1},z_t) \alpha_{z_{t-1}}(t-1) p(z_t|z_{t-1})}{p(\bm X)} \\
        = & \frac{\beta_{z_t}(t) p(\bm x_t|\bm x_1,\dots,\bm x_{t-1},z_t) \alpha_{z_{t-1}}(t-1) p(z_t|z_{t-1})}{p(\bm X)} \\
    \end{split}
\end{equation}

\subsubsection{Baum–Welch algorithm}
Now, we already have the posterior, and we proceed to the M-step of the EM algorithm.
\begin{equation}
    p(\bm X,\bm z;\theta) = p(z_1;\theta) \sbr{\prod_{t=2}^T p(z_t|z_{t-1};\theta)} \prod_{t=1}^T p(\bm x_t|
    \bm x_1,\dots,\bm x_{t-1},z_t;\theta)
\end{equation}
\begin{equation}
    \ln p(\bm X,\bm z;\theta) = \ln p(z_1;\theta) + \sum_{t=2}^T \ln p(z_t|z_{t-1};\theta) + \sum_{t=1}^T \ln p(\bm x_t|\bm x_1,\dots,\bm x_{t-1},z_t;\theta)
\end{equation}
Notice that
\begin{equation}
    \begin{split}
        Q(\theta,\theta^{\text{old}}) = & \sum_{z_{1}=1}^S \gamma_{z_1}(1)\ln p(z_1;\theta) + \sum_{t=2}^T\sum_{z_{t-1}=1}^S\sum_{z_t=1}^S \xi_{z_{t-1},z_t}(t) \ln p(z_t|z_{t-1};\theta) \\
        & + \sum_{t=1}^T\sum_{z_t=1}^S \gamma_{z_t}(t) \ln p(\bm x_t|\bm x_1,\dots,\bm x_{t-1},z_t;\theta)
    \end{split}
\end{equation}

Problems regarding the scaling factor in the forward-backward algorithm for numerical stability and the Viterbi algorithm for predicting the most probable hidden sequence are identical to the plain HMM, which can be referred to in \citep{bishop2006pattern}.

\subsection{Threshold}\label{appendix:threshold}
We show two plots of the balanced accuracy of the connection and prior matrices as a function of a threshold varying from 0 to 0.5. The plots demonstrate that, in general, the accuracy is very sensitive to the threshold.
\begin{figure}[!t]
    \centering
    \includegraphics[width=1\linewidth]{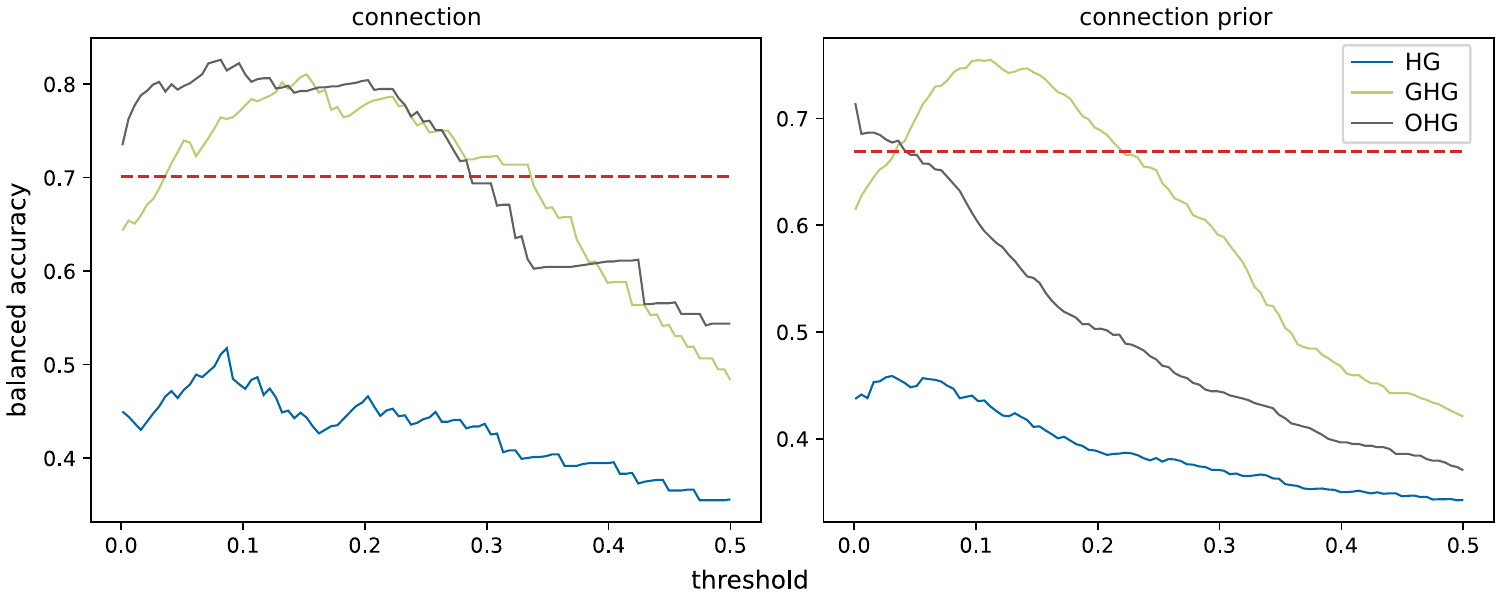}
    \caption{Using different thresholds to binarize the weight to obtain the connection. The straight dashed red line is the balanced accuracy obtained by $\bm A_0$ in OHG directly.}
    \label{fig:threshold}
\end{figure}

\subsection{Synthetic dataset}\label{appendix:synthetic}
\begin{figure}
    \centering
    \includegraphics[width=\linewidth]{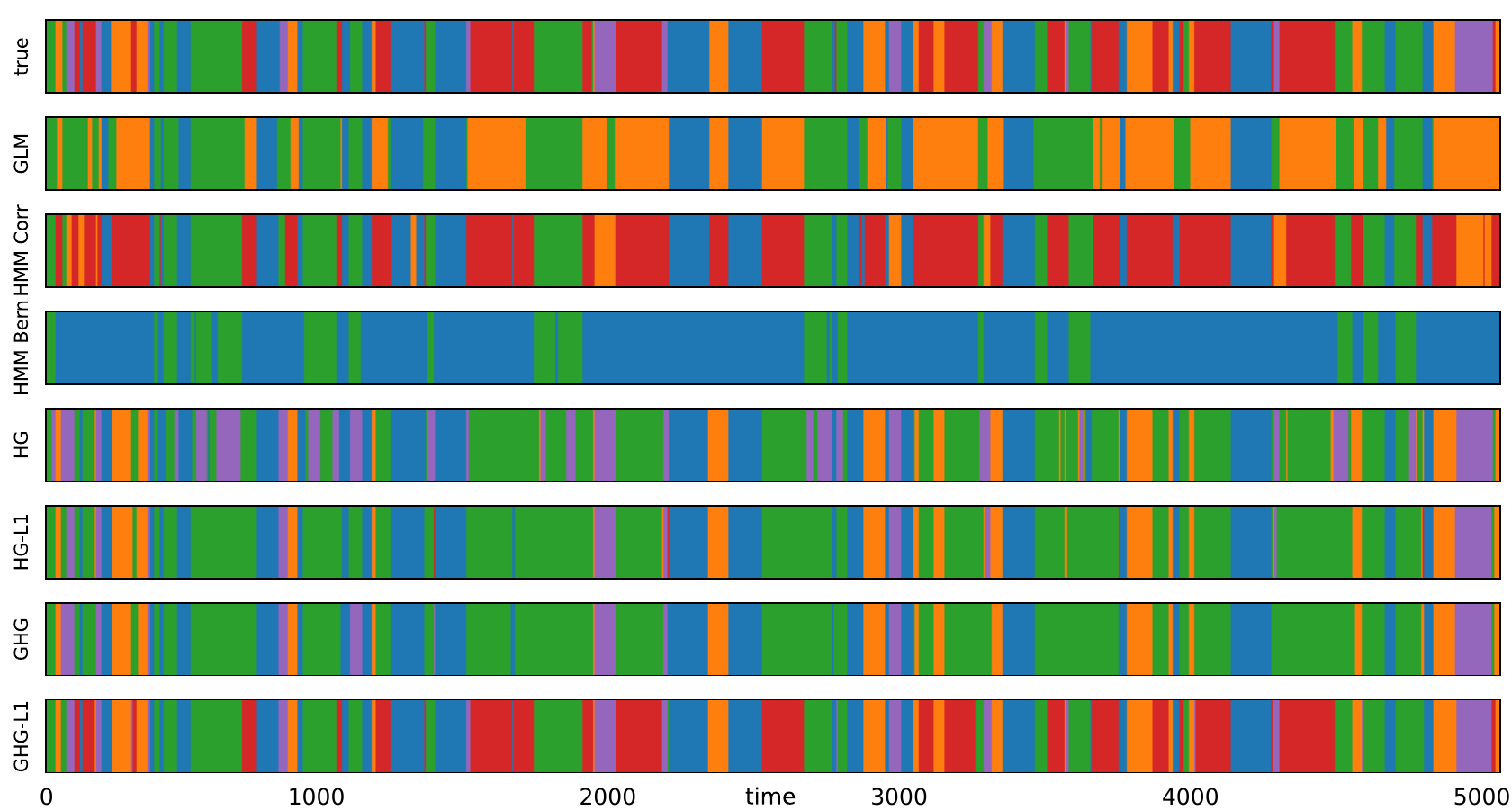}
    \caption{The state prediction of all methods applied to the one trial of the simulated spike train data. Different colors represent different states.}
    \label{fig:synthetic_states_appendix}
\end{figure}

Fig.~\ref{fig:synthetic_states_appendix} shows the state prediction of all methods on one of the synthetic spike trains.

\subsection{PFC-6 dataset}\label{appendix:pfc6}
\begin{figure}
    \centering
    \includegraphics[width=\linewidth]{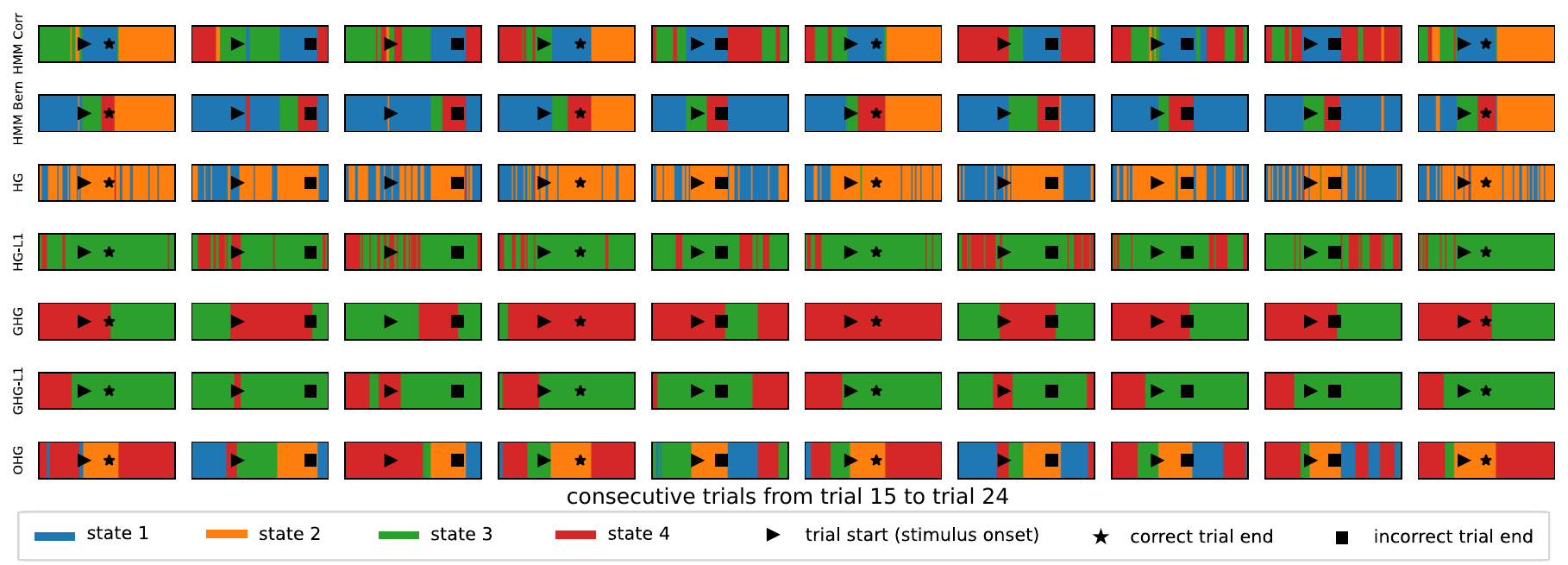}
    \caption{The state prediction of all methods applied to multiple consecutive trials of the PFC-6 spike train data.}
    \label{fig:pfc6_states_appendix}
\end{figure}

Fig.~\ref{fig:pfc6_states_appendix} shows the state prediction of all methods on trials 16-25.

\subsection{Barrel cortex data with up to 5 hidden states}\label{appendix:barrel}

\begin{figure}[!t]
    \centering
    \includegraphics[width=1\linewidth]{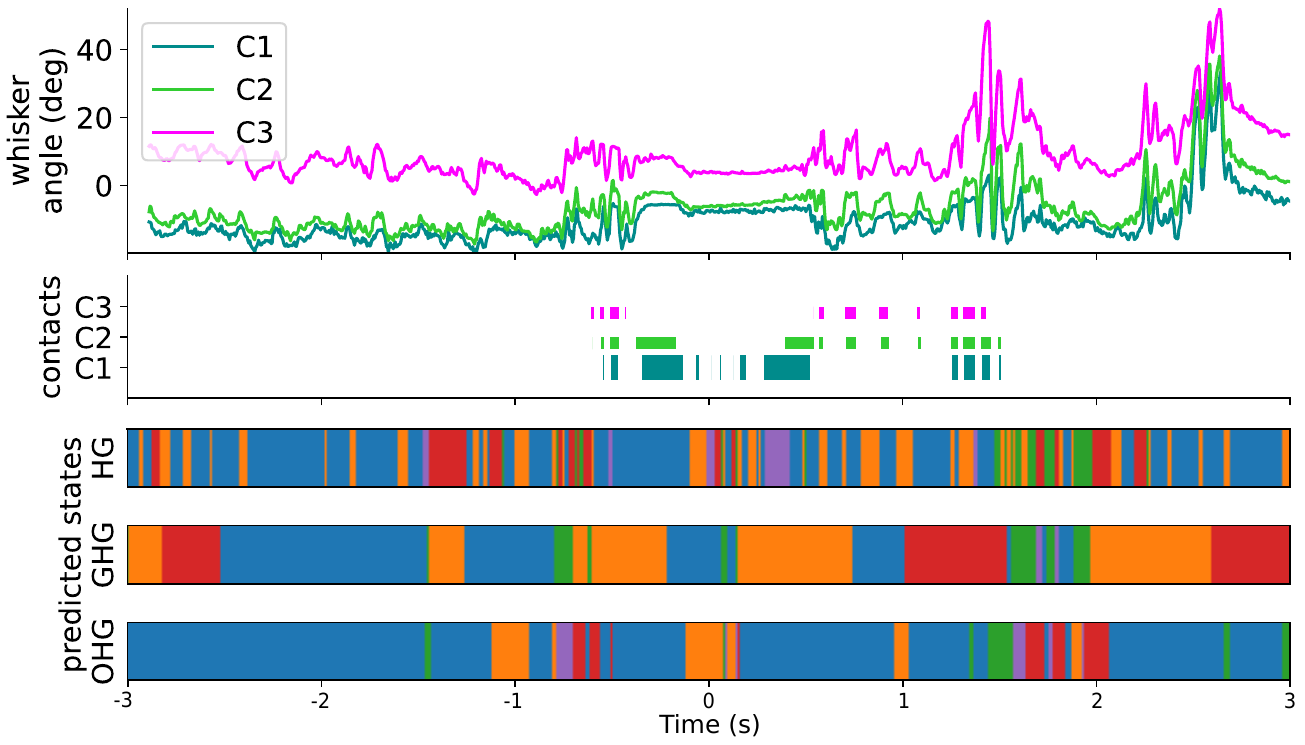}
    \caption{Example trial from barrel cortex data with $S=5$ hidden states. Different colors in the predicted states plot represent different states.}
    \label{fig:bc5states}
\end{figure}

\begin{figure}[!t]
    \centering
    \includegraphics[width=1\linewidth]{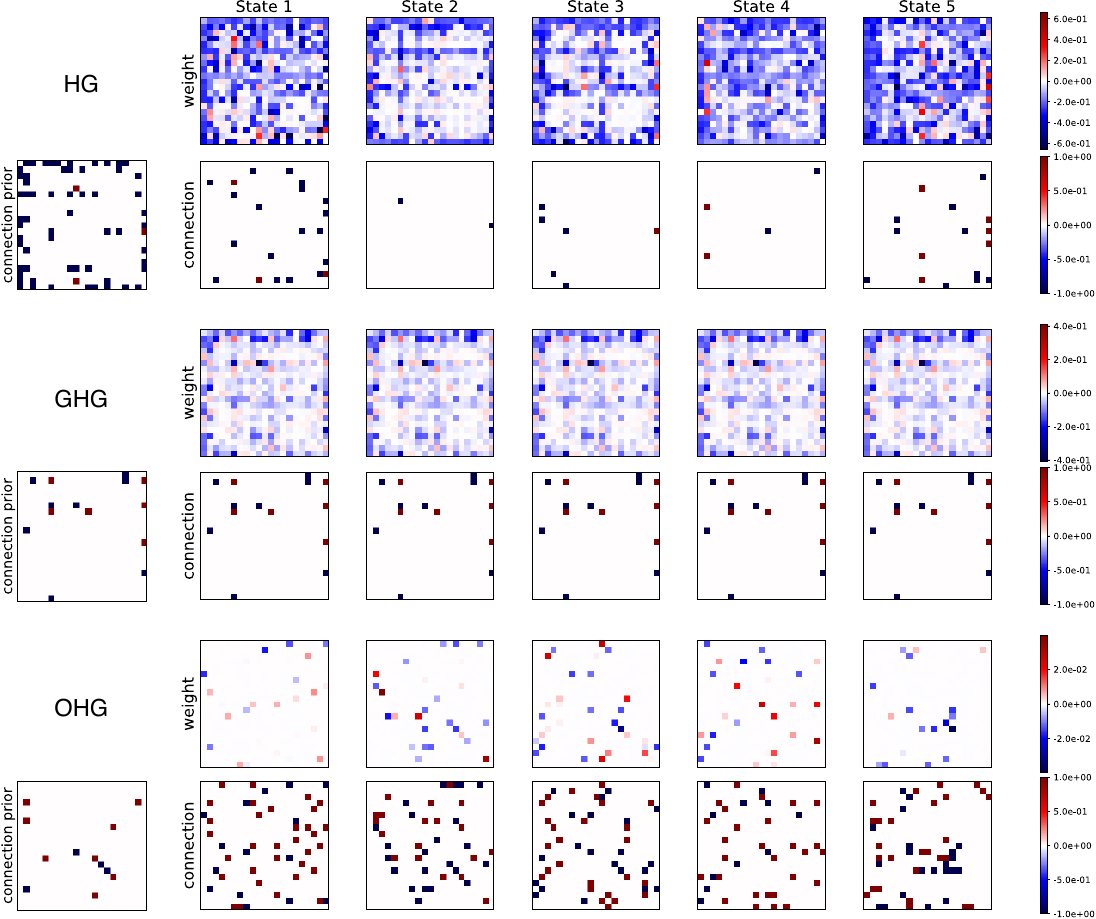}
    \caption{Weights, connection prior and connection matrices for each state of HG, GHG, and OHG models applied to the barrel cortex data session shown in Fig.~\ref{fig:bc5states}.}
    \label{fig:bc5weights}
\end{figure}

As noted in the main text, OHG exhibits increasing test log-likelihood with an increasing number of states $S$. When $S=5$, there were typically 2 or 3 dominant states predicted by OHG, with the other states being predicted only rarely across the sessions. Fig.~\ref{fig:bc5states} shows an example of a trial with $S=5$. OHG exhibits one dominant hidden state (state 5) with the other states being predicted for short intervals of duration 0.1-0.3~s, showing complex activation patterns in the vicinity of whisker contacts. The corresponding weight and connection matrices are shown in Fig.~\ref{fig:bc5weights}. Further analysis is needed to determine the significance of such states.

\end{document}